\documentclass{article}
\usepackage{spconf,amsmath,graphicx,subcaption}

\usepackage[utf8]{inputenc} 
\usepackage{url}
\usepackage{multirow}
\usepackage{multicol}
\usepackage{enumitem}
\usepackage{subcaption}
\usepackage[font=small]{caption}

\usepackage[nolist]{acronym}
\begin{acronym}
    \acro{MIR}{Music Information Retrieval}
    \acro{SDR}{Signal to Distortion Ratio} \acro{SIR}{Signal to Interference Ratio} \acro{SAR}{Signal to Artifact Ratio}
    \acro{STFT}{Short Time Fourier Transform}
    \acro{CNN}{Convolutional Neural Networks}
\end{acronym}

\title{\textsc{Singing voice separation: a study on training data}}

\name{Laure Prétet$^{\star \dagger}$ \qquad Romain Hennequin$^{\star}$ \qquad Jimena Royo-Letelier$^{\star}$ \qquad Andrea Vaglio$^{\star \dagger}$ }
\address{$^{\star}$ Deezer R\&D, Paris, France, research@deezer.com \\
    $^{\dagger}$ LTCI, Télécom ParisTech, Université Paris-Saclay, Paris, France}

\begin{document}
\ninept

\maketitle

\begin{abstract}
In the recent years, singing voice separation systems showed increased performance due to the use of supervised training. The design of training datasets is known as a crucial factor in the performance of such systems. We investigate on how the characteristics of the training dataset impacts the separation performances of state-of-the-art singing voice separation algorithms. We show that the separation quality and diversity are two important and complementary assets of a good training dataset. We also provide insights on possible transforms to perform data augmentation for this task.
\end{abstract}



\section{INTRODUCTION}
\vspace{-1mm}
Singing voice separation is the decomposition of a music recording into two tracks, the singing voice on one side, and the instrumental accompaniment on the other side. Typical applications are automatic karaoke creation, remixing, pitch tracking \cite{pollastri2002pitch}, singer identification \cite{mesaros2007singer}, and lyrics transcription \cite{mesaros2012singing}. 

This is a highly popular topic in the \ac{MIR} literature and yearly competitions such as the SiSec MUS challenge gather an increasing number of teams (24 systems evaluated in 2016, 30 in 2018).  The 2018 edition of the SiSec campaign \cite{stoter20182018} shows that the best current systems rely on supervised, deep-learning based models. In particular, \ac{CNN} seem to be especially adapted for this task. Recently, a U-Net \cite{jansson2017singing} and several DenseNet-based systems \cite{takahashi2017multi} showed impressive performance: for the first time, state-of-the-art models performed similarly to oracle systems for the instrumental part \cite{stoter20182018}.

However, despite these achievements, it is often difficult to identify what is the main success factor of these systems. Results are generally presented for a full procedure, including dataset building, data pre-processing and/or augmentation, architecture design, post-processing and sometimes a long engineering work to tune the hyperparameters of the models \cite{stoller2018wave, uhlich2017improving, jansson2017singing, stoller2017adversarial}. 

In this work, we focus on the influence of the training dataset on the performances of a state-of-the-art deep-learning based separation systems. We investigate the impact of four different aspects of these (size, separation quality, use of data augmentation techniques and use of separated sources from several instruments to estimate voice separation) by training a same baseline model while varying the training dataset. In the previous literature \cite{takahashi2018mmdenselstm, nugraha2016multichannel, luo2017deep, fan2017svsgan, stoller2017adversarial} different architectures are usually compared using the same train/test datasets, but to the best of our knowledge, there are no previous works that study particularly the influence of these datasets. As opposed to the previous works, we use one single state-of-the-art architecture and train it on different datasets in order to reveal the effect of diverse characteristics of the training data on separation performances.
We notably inspect the following aspect: data diversity and separation quality, data augmentation, and number of separated sources.

\textit{Diversity and Separation Quality.} In the literature, data scarcity is often cited as one of the main limits for building efficient and scalable supervised singing voice separation algorithms \cite{chandna2017monoaural, mimilakis2017recurrent, simpson2015deep}. Indeed, public training datasets have been regularly released (MIR-1K \cite{hsu2010improvement}, MedleyDB \cite{bittner2014medleydb}, DSD100 \cite{SiSEC17} MUSDB \cite{musdb18}) and used to compare different methods, but they are rather small, and often lack diversity. We propose here to use several datasets of different sizes and separation qualities to evaluate the benefits of training systems with larger amounts of data. These include a relatively small public database (MUSDB), a large private dataset, and a large dataset with estimated separated tracks build from Deezer's music catalog following the technique presented in \cite{humphrey2017mining}.

\textit{Data Augmentation.} A common method used to artificially increase the size of a dataset for MIR tasks is data augmentation. For instance, in singing voice detection, some data augmentation like pitch shifting or the application of random frequency filters have proven to increase performance \cite{schlueter2017_phd}. Also, in \cite{uhlich2017improving} the authors studied the use of other data augmentations (channels swapping, amplitude scaling or random chunking) with no improved results. We propose to study the influence of using several data augmentation techniques over a small sized dataset. 

\textit{Several Sources.} Finally, we study the influence of using several sources (the \emph{bass}, \emph{drums} and \emph{other} parts available in MUSDB) for estimating the \emph{instrumental} part. Indeed, when only estimating the \emph{vocal} and \emph{instrumental} parts, source separation systems tend to include in the vocals estimation residual parts from other instruments (in particular from \emph{drums}). Hence, using the additional information included in multiple sources could lead to a better modeling of the \emph{instrumental} part, and thus to a better separation.

The rest of the paper is organized as follows. In Section \ref{data}, we introduce the three datasets that we used for our experiments. In Section \ref{method}, we detail the methodology that we put in place to compare the performances on the different datasets. In Section \ref{exp}, we expose our results and discuss possible interpretations. Finally, we draw conclusions in Section \ref{conclusion}.
 
\vspace{-3mm}
\section{DATASETS}
\vspace{-1mm}

\label{data}
In this section, we present the three training datasets that we used in our experiments, along with their main characteristics. In addition to the total duration of audio, we define a \textit{quality} criterion and a \textit{diversity} criterion. The quality of the dataset reflects the quality of the source separation in the dataset's tracks: in two datasets (MUSBD and Bean), the separated tracks come from different recordings, while in the last one (Catalog), the vocal part was not available as separate track and had to be estimated. In the last case, the separated tracks being only estimates, residuals from other sources can be present in the ground truth tracks. This criterion does not account for the production quality, nor the audio quality. The diversity criterion reflects the variability of songs from which the dataset was built. It can be quantified by the number of different songs that are represented by one or more segments in the dataset. This information is summarized in Table  \ref{tab_datasets}.

\vspace{-3mm}
\subsection{MUSDB}
\vspace{-1mm}

MUSDB is the largest and most up-to-date public dataset for source separation. MUSDB is mainly composed of songs taken from DSD100 and MedleyDB datasets and was used as a reference for training and test data during the last singing voice separation campaign \cite{stoter20182018}. This dataset is composed of $150$ professionally produced songs. 0nly western music genres are present, with a vast majority of pop/rock songs, along with some hip-hop, rap and metal songs. $100$ songs belong to the training set and $50$ to the test set.

For each song, five audio files are available: the mix, and four separated tracks (\emph{drums}, \emph{bass}, \emph{vocal} and \emph{other}). The original mix can be synthesized by directly summing the tracks of the four sources. To create the \emph{instrumental} source, we add up the tracks corresponding to \emph{drums}, \emph{bass} and \emph{others}. In our experiments, we consider both the \emph{instrumental}/\emph{vocals} dataset and the 4-stems dataset.

\begin{table}[thpb]
\scriptsize
\tabcolsep=0.11cm
    \centering
    \begin{tabular}{l|c|c|c}
         & MUSDB & Catalog & Bean \\
        \hline
         Diversity & 150 songs  & 28,810 songs & 24,097 songs \\
         Quality & Separated recordings & Estimates  & Separated recordings \\
         Duration & 10 hours  & 95 hours & 79 hours \\
         Train/val/test (\%) & 53/13/33 &  97/3/0 &  85/8/7 \\  
    \end{tabular}
    \caption{Main characteristics of the three datasets.}
    \label{tab_datasets}
\end{table}

\vspace{-3mm}
\subsection{Bean}
\vspace{-1mm}

In addition to MUSDB, we use a private multi-track dataset called Bean. The Bean dataset contains a majority of pop/rock songs and includes both vocal and instrumental tracks as separated recordings. Among the 24,097 available songs in this dataset, 21,597 were used for training, 2,000 for validation and the 500 remaining for test.

In total, the Bean dataset represents 5,679 different artists. We made the train/validation/test split in such a way that an artist cannot appear simultaneously in two parts of the split, as in MUSDB. This is an important precaution to ensure that the separation system will not overfit on the artists, an issue often raised in \ac{MIR} \cite{flexer2007closer}. 
We performed genre statistics on Bean, as presented in Figure \ref{fig:genre_distribution} (green histogram). The genre distribution in Bean is mainly dominated by Pop and Rock songs, which is quite similar to MUSDB.

\vspace{-3mm}
\subsection{Catalog}
\vspace{-1mm}

To build this dataset, we took inspiration from \cite{humphrey2017mining}, where a method is presented to build a dataset based on a music streaming catalog. We adapted this method to build a dataset from Deezer's catalog, by exploiting the \textit{instrumental versions} that are released by some artists along with the original songs.

The first step is to find all possible \textit{instrumental/mix} track pairs within the catalog. This matching is done using metadata and audio fingerprinting. 
Then, a few filtering and homogenization operations are performed: A pair is removed if both tracks have a duration difference greater than 2 seconds. Songs longer than 5 minutes are filtered out. Then, tracks within a pair are temporally re-aligned using autocorrelation. Finally, the loudness of both tracks is equalized.

To produce a triplet \textit{(mix, instrumental, vocals)} from the pair \textit{(mix, instrumental)}, we perform a half-wave rectified difference of both spectrograms. Eventually, 28,810 triplets were created. We split them into a training and a validation dataset, making sure that a given artist cannot appear simultaneously in both parts of the split. We refer this dataset as \emph{Catalog A}.

\begin{figure}[thpb]
       \centering
       \includegraphics[width=\columnwidth]{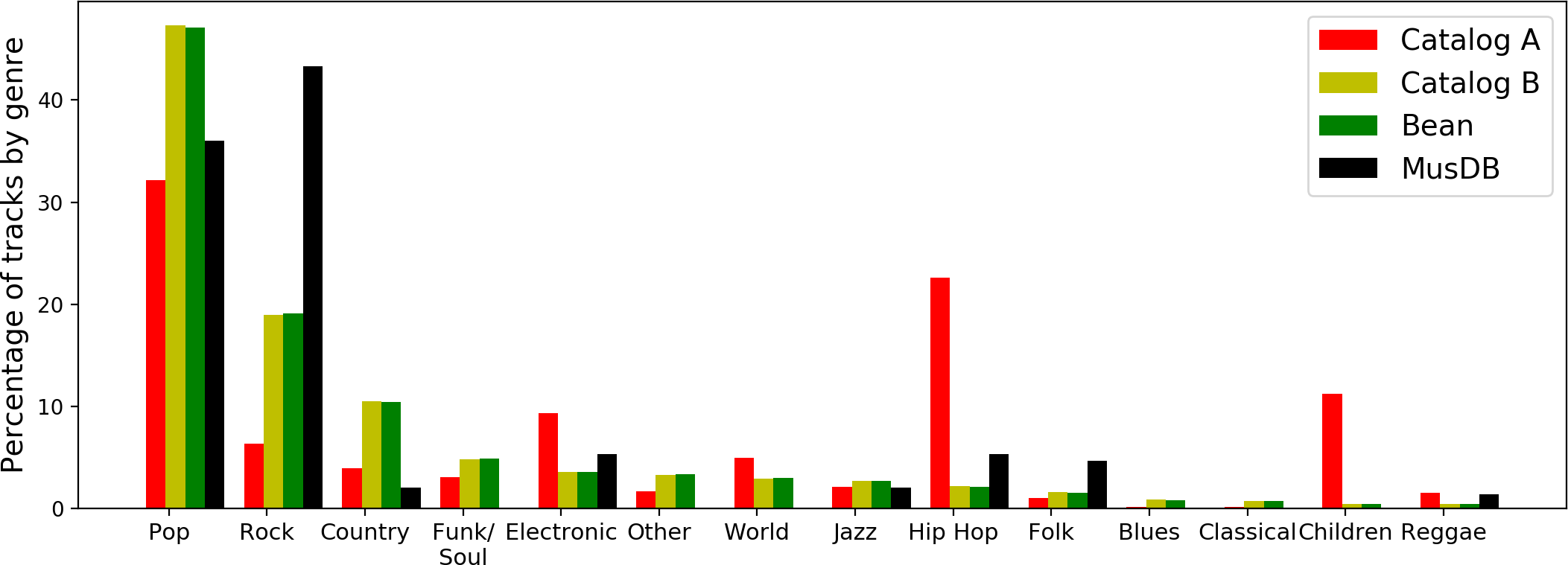}
       \caption{Genre distribution for Bean, Catalog and MUSDB datasets.}
       \label{fig:genre_distribution}
 \end{figure}

Using metadata, we noticed an important genre bias towards kids music and hip-hop in this dataset compared to the genre distribution in Bean (and consequently in MUSDB), as represented in Figure \ref{fig:genre_distribution}.
To overcome this issue, we built a second dataset by re-balancing the representation of each genre in a way that the final distribution matches the one of Bean. We refer to this dataset as \emph{Catalog B}.

Even though Catalog benefits from a very large volume compared to MUSDB, we must keep in mind that it was not professionally produced for separation purposes and is necessarily of a lower quality. The two main issues that we found in the dataset are:
\begin{itemize}[leftmargin=4mm, parsep=0cm, itemsep=0cm, topsep=0cm]
\item The half-wave rectified difference between the mix and the instrumental does not correspond exactly to the vocal part. This is because this operation is performed on magnitude spectrograms, for which source additivity is not ensured. Besides, the smallest misalignment between both tracks can produce instrumental residuals in the vocals. An informal listening test on a small subset (40 tracks) reveals that this happens in almost 50\% of the tracks.
\item If the metadata matching is not perfect, there may be songs with no singing voice in the mix. In this case, the \emph{vocals} part is only a residual noise. Reversely, some \emph{instrumental} tracks contain choirs. These cases are difficult to detect by automatic systems.
\end{itemize}

Accordingly, we may say that the Catalog database forms a large amount of weakly labeled training data. The instrumental part is professionally-produced, while the vocals are only estimates.


\vspace{-3mm}
\section{METHODOLOGY}
\vspace{-1mm}

\label{method}

\vspace{-1mm}
\subsection{Network architecture}
\vspace{-1mm}

In this paper, we focus on deep neural networks to perform the separation. The baseline model that we chose is the U-Net, as proposed in \cite{jansson2017singing}. After some pilot experiments with other architectures (the DenseNet and MMDenseNet from \cite{takahashi2017multi}), we selected the U-Net, which could train in a reasonable amount of time even on large datasets. This architecture showed state-of-the-art results on the DSD100 dataset \cite{jansson2017singing} and in the last SiSeC campaign \cite{stoter20182018}. It is also a simple, general architecture that can be applied in a variety of domains \cite{ronneberger2015u}.

The U-Net shares the same architecture as a convolutional auto-encoder with extra skip-connections that bring back detailed information lost during the encoding stage to the decoding stage. 
It has 5 strided 2D convolution layers in the encoder and 5 strided 2D deconvolution layers in the decoder.

The main modification compared to \cite{jansson2017singing} was to integrate stereo processing: we used 3D tensors (channels, time
steps, frequency bins) as input and output of the network. The other layers were not modified.

\vspace{-3mm}
\subsection{Data preparation}
\vspace{-1mm}

In the original datasets, all songs are stereo and sampled at $44100$Hz. To reduce computational cost, we resample them to $22050$Hz . We split all songs into segments of $11.88$ seconds. For Catalog and Bean, we randomly select one segment from each song in the training and validation sets, avoiding the intro (first $20$s) and the outro (last $20$s), where vocals are often missing. We also constructed a second test dataset using $500$ tracks from Bean, from which we were able to extract 1,900 segments.
We made sure to balance its genre distribution over the $10$ most represented genres of Figure \ref{fig:genre_distribution}. The final split proportions can be seen in Table \ref{tab_datasets}.

Similarly to \cite{jansson2017singing}, we used \ac{STFT} as input and output features for our network. The window size is $2048$ and the step size is $512$. We chose these settings such that after removing the highest frequency band, the dimensions of the spectrograms are a power of $2$: (channels, time steps, frequency bins) = ($2$, $512$, $1024$). This is necessary, because the network architecture that we use reduces the dimensions of the spectrograms by a factor which is a power of two.

\vspace{-3mm}
\subsection{Training}
\vspace{-2mm}

For each source (\emph{vocals} and \emph{instrumental}), we trained a U-Net to ouptut the corresponding magnitude spectrogram from the magnitude spectrogram of the mixture. We trained each network for $500$ epochs using Keras with Tensorflow backend. We define one epoch as $800$ gradient descent steps. To limit overfitting, we use the validation split of each dataset for early stopping. The training loss is the $L_1$ norm of the difference between the target spectrogram and the masked output spectrogram, as described in \cite{jansson2017singing}. The optimizer is ADAM and the learning rate is $0.0001$. The batch size is set to $1$ after a short grid search.

\vspace{-3mm}
\subsection{Reconstruction}
\vspace{-2mm}

Once the training is finished, we perform an inference pass on the test dataset, equally cut into $11.88$ second segments. The complex spectrograms of each source are reconstructed by computing a ratio mask from both estimates and applying it to the original mixture spectrogram. This way, the output phase is that of the mixture. The ratio mask of a source is obtained by dividing the spectrogram estimate of a source (output of the corresponding U-Net) by the sum of both the estimates. For the particular case of $4$-stems separation, the \emph{instrumental} spectrogram estimates is obtained by summing the spectrogram estimates of the $3$ non-vocals stems.
The \ac{STFT} are inverted and full songs are reconstructed by simply concatenating the different segments. The audio is finally upsampled back to $44100$Hz.

\vspace{-3mm}
\subsection{Evaluation}
\vspace{-2mm}

We use the Museval \cite{musdb18} toolbox to compute the standard source separation measures: \ac{SDR}, \ac{SIR} and \ac{SAR}. We aggregate these metrics using a median over all $1$-second frames to keep one single metric per song and per source, as in \cite{stoter20182018}. We run the evaluation process on both the MUSDB and Bean test datasets.

To compare the performance of the different methods, we also conducted a paired Student $t$-test on the per songs metrics. 
This step was motivated by the observation that the variance was high in the metric distributions, making it sometimes difficult to assess whether a method performed significantly better than another one or not.
Even though two methods may produce very similar distributions of the metrics, these metrics may vary in a dependent way (e.g. with a small but constant difference). The paired $t$-test helps revealing this phenomenon.

\vspace{-3mm}
\section{EXPERIMENTS AND RESULTS}
\vspace{-1mm}

\label{exp}

\vspace{-2mm}
\subsection{Data augmentation}
\vspace{-1mm}

When training on a small dataset like MUSDB, data augmentation is regularly cited as a way to improve separation performances \cite{uhlich2017improving}. In this experiment, we try to figure out to what extent data augmentation can improve separation performances. For selecting data transformation to be performed, we took inspiration from \cite{schlueter2017_phd}, in which the author uses a set of transformations on the spectrograms and tests the effect on a singing voice detection task. We set up a similar set of experiments to evaluate the impact of various forms of data augmentation on separation results. We adapted the transforms proposed by Schülter (pitch shifting, time stretching, loudness modification and filtering) for source separation and added channel swapping (following \cite{uhlich2017improving}) and source remixing. The specificity of data augmentation in the context of source separation is that both the target and the inputs must be processed with the exact same transformation.
Here is the detail of the various transformations we used:

\textbf{Channel swapping [Swap]:} The left and right channels are swapped with a probability of 0.5.

\textbf{Time stretching [Stretch]:} We linearly scale the time axis of the spectrograms by a factor $\beta_{stretch}$ and keep the central part. $\beta_{stretch}$ is drawn randomly from a uniform distribution between 0.7 and 1.3 ($\pm$ 30\%) for each sample. Note that this is an approximation compared to an actual modification of the speed of the audio.

\textbf{Pitch shifting [Shift]:} We linearly scale the frequency axis of the spectrograms by a factor $\beta_{shift}$ and keep the bottom part, such that the lowest frequency band stays aligned with 0 Hz. $\beta_{shift}$ is drawn randomly from a uniform distribution between 0.7 and 1.3 ($\pm$ 30\%) for each sample. Note that this is an approximation compared to an actual pitch shifting of the audio.

\textbf{Remixing [Remix]:} We remix the \emph{instrumental} and \emph{vocals} part with random loudness coefficients, drawn uniformly on a logarithmic scale between $-9$dB and $+9$dB.

\textbf{Inverse Gaussian filtering [Filter]:} We apply to each sample a filter with a frequency response of $f(s) = 1- e^{-(s - \mu)^2/2\sigma^2}$ with $\mu$ randomly chosen on a linear scale from 0 to $4410$Hz and $\sigma$ randomly chosen on a linear scale from $500$Hz to $1000$Hz.

\textbf{Loudness scaling [Scale]:} we multiply all the coefficients of the spectrograms by a factor $\beta_{scale}$. $\beta_{scale}$ is drawn uniformly on a logarithmic scale between $-10$dB and $+10$dB.

\textbf{Combined:} We perform simultaneously the channel swapping, pitch shifting, time stretching and remixing data augmentations.

\begin{figure}
    \centering
    ~ 
    \begin{subfigure}[b]{0.3\columnwidth}
        \includegraphics[scale=0.4]{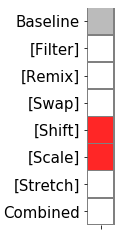}
        \caption{Voice}
        \label{fig:augmentation_voice}
    \end{subfigure}
    ~ 
    \begin{subfigure}[b]{0.3\columnwidth}
        \includegraphics[scale=0.4]{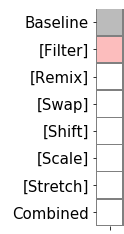}
        \caption{Instruments}
        \label{fig:augmentation_instr}
    \end{subfigure}
    \begin{subfigure}[b]{0.1\columnwidth}
        \includegraphics[scale=0.18]{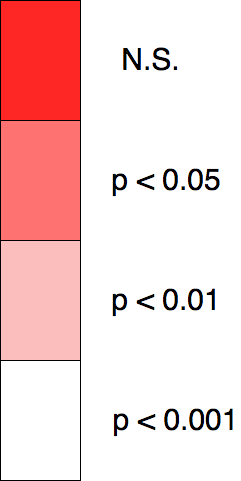}
        \vspace{0.3cm}
    \end{subfigure}
    \caption{Data augmentation experiment: Results of the Student's paired $t$-test for the \ac{SDR} on the MUSDB Test dataset.}
    \label{fig:augmentation}
\end{figure}
 
Median source separation metrics (\ac{SDR}, \ac{SAR}, \ac{SIR}) are reported in Table \ref{tab:res_augmentation}. To get an idea of the significance of the metric differences, we performed a paired Student $t$-test between data augmented training and the not data augmented baseline: we report $p$-values for this test applied to \ac{SDR} on the MUSDB test set in Figure \ref{fig:augmentation}.

Table \ref{tab:res_augmentation} shows that data augmentation may have a positive impact on separation metrics in some case: notably on the Bean dataset, channel swapping, pitch shifting and time-stretching seems to quite consistently improve most of the metrics. However it must be noted that even when the improvement is statistically significant for the test we performed, the improvement is very limited and hardly exceeds $0.2$dB in \ac{SDR}, which is very low and might not even be audible. Thus, the various data augmentation types we tested seem to have quite a low impact on separation results while being commonly used in the literature.

\begin{table}[thpb]
\scriptsize
  \begin{tabular}{|l|l|l|l|l|l|l|l|}
    \hline
    & & \multicolumn{3}{c|}{Voice} &
      \multicolumn{3}{c|}{Instruments} \\
    \hline
    Test & Transform & SDR & SIR & SAR & SDR & SIR & SAR \\
    \hline
    \multirow{4}{*}{MUSDB} &\textit{Baseline}&\textit{4.32}&\textit{12.62}&\textit{4.1}&\textit{10.65}&\textit{13.46}&\textit{11.51} \\
 &[Filter]&3.9&\textbf{13.35}&3.33&10.27&12.57&11.66 \\
 &[Remix]&3.75&12.89&3.6&10.45&11.81&\textbf{12.05} \\
 &[Swap]&4.37&\textbf{13.01}&4.08&\textbf{10.69}&13.08&\textbf{11.74} \\
 &[Shift]&4.0&\textbf{15.3}&3.5&10.58&12.46&\textbf{12.11} \\
 &[Scale]&4.05&12.6&3.64&10.68&12.38&\textbf{11.85} \\ 
 &[Stretch]&4.19&\textbf{13.44}&3.57&10.96&12.76&\textbf{12.09} \\ 
 &Combined&3.76&\textbf{13.86}&3.3&10.48&12.35&\textbf{11.72} \\
    \hline
    \multirow{4}{*}{Bean} &\textit{Baseline}&\textit{5.91}&\textit{9.23}&\textit{5.73}&\textit{9.33}&\textit{12.43}&10.9 \\ 
 &[Filter]&5.58&\textbf{10.8}&5.2&9.18&11.53&10.75 \\ 
 &[Remix]&5.7&\textbf{10.18}&5.44&9.43&11.1&\textbf{11.4} \\ 
 &[Swap]&\textbf{5.98}&\textbf{9.94}&\textbf{5.83}&\textbf{9.5}&12.25&\textbf{11.24 }\\ 
 &[Shift]&\textbf{6.06}&\textbf{11.53}&5.82&\textbf{9.57}&11.67&\textbf{11.63} \\ 
 &[Scale]&5.87&\textbf{9.55}&5.66&9.42&11.71&\textbf{11.32}  \\ 
 &[Stretch]&\textbf{6.12}&\textbf{10.68}&\textbf{5.94}&\textbf{9.64}&12.18&\textbf{11.35} \\ 
 &Combined&5.98&\textbf{11.45}&\textbf{5.99}&9.4&11.1&\textbf{11.07} \\ 
    \hline
  \end{tabular}
  \caption{Data augmentation experiment: Results of the U-Net trained on MUSDB with data augmentation. In bold are the results that significantly improve over the baseline (p $<$ 0.001).}
  \label{tab:res_augmentation}
\end{table}

\vspace{-5mm}
\subsection{Impact of the training dataset}
\vspace{-1mm}

\begin{figure}
    \centering
    \begin{subfigure}[b]{0.4\columnwidth}
        \includegraphics[width=\textwidth]{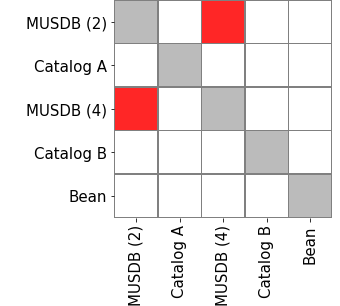}
        \caption{Voice}
        \label{fig:datasets_voice}
    \end{subfigure}
    \begin{subfigure}[b]{0.4\columnwidth}
        \includegraphics[width=\textwidth]{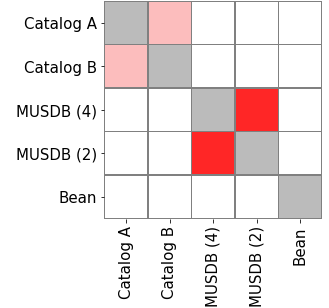}
        \caption{Instruments}
        \label{fig:datasets_instr}
    \end{subfigure}
    \begin{subfigure}[b]{0.1\columnwidth}
        \includegraphics[scale=0.14]{Figures/legend.png}
            \vspace{1cm}
    \end{subfigure}
    \caption{Training dataset comparison experiment: Results of the Student's paired $t$-test for the SDR on the MUSDB Test dataset. SDR increases from top left to bottom right.}
    \label{fig:datasets}
\end{figure}

In this experiment, we evaluate the impact of the training dataset on the performances of the selected separation system. The system is trained with the $5$ datasets presented in Section \ref{data}: \emph{Catalog A}, \emph{Catalog B}, Bean, MUSDB with two stems (\emph{accompaniment} and \emph{vocals}) and MUSDB with four stems (\emph{vocals}, \emph{drums}, \emph{bass} and \emph{other}). After training the system on each dataset, we evaluate its performances on the two test datasets: MUSDB and Bean.
Medians over all tracks of source separation metrics are reported in Table \ref{tab:res_datasets} and $p$-values for the paired Student $t$-test between SDR obtained on the MUSDB test dataset are reported in Figure \ref{fig:datasets}. 

As expected, training on the Bean dataset yields the highest scores for most metrics on both the vocals and the accompaniment parts and on both test datasets. It is worth noting that the SDR values on the \emph{vocals} part for the system trained on Bean are higher than the ones for all other systems by more than $1$dB on the MUSDB test set and $1.5$dB on the Bean test set, which is quite important (and is perceptually very noticeable). This confirms that having large datasets with clean separated tracks is a good way of improving performances of source separation systems.
More surprisingly, all other training datasets provide quite similar performances from one to another. In particular, training on $4$ stems instead of $2$ did not improve significantly the metrics on MUSDB: then on this particular setup, adding extra information to help modelling the accompaniment spectrogram actually did not result in improved performance.

We also notice that training the system with both \emph{Catalog} datasets has a very limited impact on the separation performances. 
Compared to MUSDB alone, it yields in higher SAR, but lower SIR, resulting in a similar SDR. The effect is particularly visible on the vocals. This makes sense with the way the Catalog training dataset was built: the recordings are professionally produced, so the mixture quality is good, but significant leaks remain in the vocal target. Moreover, training with \emph{Catalog A} or \emph{Catalog B} seems to provide very similar results, which means that the difference of genre distribution between \emph{Catalog A} and \emph{Bean} is not responsible for the high differences of performance and the actual reason for lower performance is probably the lower quality of the separated tracks of the dataset.

Hence, training a system on a large and diverse dataset with low quality semi-automatically obtained sources seems to have a very limited impact on the performance metrics compared to using a large clean dataset such as Bean. This comes in contradiction to what was suggested in \cite{jansson2017singing}, where the impact of the size of the dataset was assumed to be important (even though this aspect was not tested with all other factor being fixed).

\begin{table}[thpb]
\scriptsize
\tabcolsep=0.11cm
  \begin{tabular}{|l|l|l|l|l|l|l|l|}
    \hline
    & & \multicolumn{3}{c|}{Voice} &
      \multicolumn{3}{c|}{Instruments} \\
    \hline
    Test & Train & SDR & SIR & SAR & SDR & SIR & SAR \\
    \hline
    \multirow{4}{*}{MUSDB}  & MUSDB (2 stems) & 4.32 & 12.62 & 4.1 & 10.65 & 13.46 & 11.51 \\
 &MUSDB (4 stems)&4.44&12.26&4.2&10.61&13.7&11.48 \\
 &\emph{Catalog A}&4.2&7.6&\textbf{7.44}&10.47&12.84&12.03 \\
 &\emph{Catalog B}&4.34&8.04&7.05&10.6&12.8&12.12 \\
 &Bean&\textbf{5.71}&\textbf{14.82}&5.19&\textbf{11.99}&\textbf{16.04}&\textbf{12.21} \\
    \hline
    \multirow{4}{*}{Bean} &MUSDB (2 stems) & 5.91 & 9.23 & 5.73 & 9.33 & 12.43 & 10.9 \\ 
 &MUSDB (4 stems)&5.88&8.56&5.71&9.3&12.87&10.92 \\
 &\emph{Catalog A}&5.85&7.26&7.16&9.56&11.68&12.3 \\
 &\emph{Catalog B}&6.05&7.62&6.79&9.74&11.85&\textbf{12.42} \\
 &Bean&\textbf{7.67}&\textbf{12.33}&\textbf{7.51}&\textbf{11.09}&\textbf{15.35}&12.17 \\
    \hline
  \end{tabular}
  \caption{Training dataset comparison experiment: Results of the U-Net system trained on the 5 different datasets. The best results on each test dataset are displayed in bold.}
  \label{tab:res_datasets}
\end{table}

\vspace{-4mm}
\section{CONCLUSION}
\vspace{-1mm}

\label{conclusion}

In this study, we consider what aspects of training datasets have an impact on separation performances for a particular state-of-the-art source separation system (U-Net).
In this setup, we showed that data augmentation, while quite frequently used in the literature, has a very limited impact on the separation results when performed on a small training dataset. We also showed that the extra information brought by having access to more sources than needed for performing the separation task ($4$ stems instead of \emph{vocals} and \emph{accompaniment} only) does not improve the system performances. Besides, we showed that, as opposed to what was assumed in the literature, a large dataset with semi-automatically obtained vocal sources does not help much the studied system compared to a smaller dataset with separately recorded sources. At last, we confirmed a common belief that having a large dataset with clean separated sources improves significantly separation results over a small one.

In future works, we may try to generalize these results to other state-of-the-art sources separation systems. Moreover, we focused on objective source separation metrics that are known to poorly account for perceptual differences between system. Then, assessing the impact of data with a stronger focus on the perceptual impact would be a relevant continuation of this work.


\bibliographystyle{IEEEbib}
\bibliography{camera_ready}

\end{document}